\documentclass[aps,prl,twocolumn,groupedaddress,showpacs]{revtex4}

\usepackage{graphicx}
\usepackage{dcolumn}
\usepackage{bm}

\begin{document}

\title{Three-Pion Hanbury-Brown-Twiss Correlations in 
Relativistic Heavy-Ion Collisions from the STAR Experiment}

\author{J.~Adams$^3$, C.~Adler$^{11}$, Z.~Ahammed$^{24}$,
C.~Allgower$^{12}$,
J.~Amonett$^{14}$, B.D.~Anderson$^{14}$, M.~Anderson$^5$, 
D.~Arkhipkin$^{10}$, G.S.~Averichev$^{9}$, 
J.~Balewski$^{12}$
% ...
\\
 (STAR Collaboration)}

\affiliation{
% ...
$^{3}$University of Birmingham, Birmingham, United Kingdom\\
% ...
$^{5}$University of California, Davis, California 95616\\
% ...
$^{9}$Laboratory for High Energy (JINR), Dubna, Russia\\
$^{10}$Particle Physics Laboratory (JINR), Dubna, Russia\\
$^{11}$University of Frankfurt, Frankfurt, Germany\\
$^{12}$Indiana University, Bloomington, Indiana 47408\\
% ...
$^{14}$Kent State University, Kent, Ohio 44242\\
% ...
$^{24}$Purdue University, West Lafayette, Indiana 47907\\
% ...
}
\date{\today{}}

\begin{abstract}
Data from the first physics run at the Relativistic Heavy-Ion Collider
at Brookhaven National Laboratory, Au+Au collisions
at $\sqrt{s_{NN}}=130$ GeV, have been analyzed by the STAR Collaboration
using three-pion correlations with charged pions to study whether pions are
emitted independently at freezeout. We have made a high-statistics
measurement of the three-pion correlation function and calculated
the normalized three-particle correlator to obtain a quantitative
measurement of the degree of chaoticity of the pion source.
It is found that the degree of chaoticity seems to increase with
increasing particle multiplicity.
\end{abstract}

% insert suggested PACS numbers in braces on next line
% insert suggested keywords - APS authors don't need to do this
%\keywords{}

\maketitle
Two-pion Hanbury Brown and Twiss (HBT) interferometry in principle
provides a means of extracting the space-time evolution of the pion
source produced at kinematic freeze-out in 
relativistic heavy-ion collisions
\cite{Gyulassy,Report}. An underlying assumption of this method is
that pions are produced from a completely chaotic source, i.e. a source
in which the hadronized pions are created with random quantum particle
production phases. In applications of two-pion HBT the validity of
this assumption is usually tested by extracting 
the {}``\( \lambda  \)-parameter'' which in a simple picture is 
unity for a fully chaotic
source and zero for a fully coherent source \cite{Gyulassy,Report}.
However, this parameter also depends on many other factors, such as
contamination from other particles in the pion sample, unresolvable
contributions from the decay of long-lived resonances and unstable
particles (\( \omega ,\, \eta ,\, \eta' ,\, K^{0},\, \Lambda  \),
etc.), and inaccurate Coulomb corrections \cite{Report}.

A better determination of the source chaoticity is possible by using
three-particle correlations. Normalizing the three-pion correlation
function appropriately by the two-pion correlator, the effects from
particle misidentification and decay contributions can be
removed \cite{HZ}, thereby isolating possible coherence effects
in the particle emission process. The resulting three-pion correlator
\( r_{3} \) provides the means of extracting the degree of source
chaoticity by examining its value in the limit of 
zero relative momentum. Recent
measurements at the CERN SPS from experiments NA44 and WA98
have focused on extracting \( r_{3} \) from three-pion correlations
\cite{NA44,WA98}. While these studies have produced results which
are consistent with a chaotic source for Pb+Pb collisions
(\( \sqrt{s_{NN}}=17 \) GeV), NA44 in
particular has shown for S+Pb collisions (\( \sqrt{s_{NN}}=20 \) GeV) 
a result which does not appear
to be consistent with the chaotic assumption. All of these prior results
suffer from low statistics which limits their significance. We
present here using charged pions the first high-statistics heavy-ion study 
of three-pion correlations,
resulting in the first accurate measurement of the degree of chaoticity
in Au+Au collisions at the Relativistic Heavy Ion Collider (RHIC). 
Note that a similar study has recently
been carried out for CERN LEP $e^{+}e^{-}$ collisions \cite{LEP}
which reports a fully chaotic source.

The present three-pion correlation study by the STAR experiment at
RHIC supplements the published two-pion correlation data from Au+Au
collisions at \( \sqrt{s_{NN}}=130 \) GeV \cite{STARHBT}. A summary
of the three-pion results will be presented for two multiplicity
classes. By looking at collision classes with different multiplicities
we can vary the impact parameter, and thus the number of initially
colliding nucleons, and study the effect of the size of the colliding
system on the source chaoticity. We discuss the method of normalization
of the correlation function and its extrapolation to vanishing relative
momentum in order to extract the source chaoticity; we estimate the
various systematic uncertainties associated with these procedures.

Before presenting our experimental results, we first outline the formalism
which guided our analysis (for details see Ref.~\cite{HZ,NS} and references
therein). The measured observable is the normalized three-pion correlator:
\begin{widetext}

\begin{equation}
\label{CosPhi}
r_{3}\left( Q_{3}\right) = \frac{\left( C_{3}\left( Q_{3}\right) -1\right) 
-\left( C_{2}\left( Q_{12}\right) -1\right) -\left( C_{2}\left( Q_{23}
\right) -1\right) -\left( C_{2}\left( Q_{31}\right) -1\right) }{\sqrt{
\left( C_{2}\left( Q_{12}\right) -1\right) \left( C_{2}\left( Q_{23}
\right) -1\right) \left( C_{2}\left( Q_{31}\right) -1\right) }}
\end{equation}

\end{widetext}Here \( Q_{3}=\sqrt{Q_{12}^{2}+Q_{23}^{2}+Q_{31}^{2}} \)
and \( Q_{ij}=\sqrt{-(p_{i}{-}p_{j})^{2}} \) are the standard invariant
relative momenta \cite{NA44,WA98} which can be computed for each
pion triplet from the three measured momenta \( ({\textbf {p}}_{1},
{\textbf {p}}_{2},{\textbf {p}}_{3}) \).
\( C_{2}\left( p_{i},p_{j}\right) =\frac{P_{2}\left( p_{i},p_{j}\right) }
{P_{1}\left( p_{i}\right) P_{1}\left( p_{j}\right) } = C_2(Q_{ij}) \)
and \( C_{3}\left( p_{1},p_{2},p_{3}\right) =\frac{P_{3}\left( p_{1},
p_{2},p_{3}\right) }{P_{1}\left( p_{1}\right) P_{1}\left( p_{2}\right) 
P_{1}\left( p_{3}\right) } = C_3(Q_3) \),
where \( P \) represents the momentum probability distribution. In
Ref.~\cite{HZ} the ratio \( r_{3} \) is defined in terms of functions
which depend on all 9 components of \( ({\textbf {p}}_{1},
{\textbf {p}}_{2},{\textbf {p}}_{3}) \);
however, limited statistics even in our high-statistics sample requires
a projection of both the numerator and denominator onto a single momentum
variable, \( Q_{3} \).

For fully chaotic sources \( r_{3}/2 \) approaches unity as all
relative momenta (and thus \( Q_{3} \)) go to zero. If the source
is partially coherent, a relationship can be established \cite{HZ}
between the
limiting value of the three-pion correlator at \( Q_{3}=0 \) and
\( \varepsilon  \),
the fraction of pions which are emitted chaotically from the pion source 
(\( 0{\, \leq \, }\varepsilon {\, \leq \, }1 \)):
\begin{equation}
\label{Chaotic Fraction}
\frac{1}{2}\, r_{3}\left( Q_{3}{=}0\right) =\sqrt{\varepsilon }\, 
\frac{3-2\varepsilon }{(2{-}\varepsilon )^{3/2}}.
\end{equation}
\( \varepsilon  \) gives an upper limit on the value of the two-pion
$\lambda$-parameter, which is sensitive to the fraction of 
coherent pairs in a sample, 
i.e. $\lambda=\varepsilon(2-\varepsilon)$ assuming no other effects
on $\lambda$ such as long-lived resonances\cite{Report}.
Eq. (2) is not affected by the projection onto a single relative
momentum variable. To exploit it and extract the degree of chaoticity
\( \varepsilon  \), the measured data for \( r_{3} \) must, however,
be extrapolated from finite \( Q_{3} \) to \( Q_{3}{\, =\, }0 \).

Similar to the two-boson correlation function, the three-boson correlation
function \( C_{3}\left( Q_{3}\right)  \) is calculated from the data
by taking the ratio \( \frac{A\left( Q_{3}\right) }{B\left( Q_{3}\right) } \)
and normalizing it to unity at large \( Q_{3} \). 
Here \( A\left( Q_{3}\right) =\frac{dN}{dQ_{3}} \)
is the three-pion distribution as a function of the invariant three-pion
relative momentum, integrated over the total momentum of the pion
triplet as well as all other relative momentum components. It is obtained
by taking three pions from a single event, calculating \( Q_{3} \),
and binning the results in a histogram. \( B\left( Q_{3}\right)  \)
is the analogous mixed event distribution which is computed by taking
a single pion from each of three separate events. 
Because of the zero in the
denominator of the normalized three-pion correlator \( r_{3} \) at
large \( Q_{ij} \), the particular method of normalization of \( C_{2} \)
and \( C_{3} \) can have a strong effect on the calculation. The propagation
of statistical errors through the \( r_{3} \) functions, however,
accounts for these effects completely. In fact, it is only with the
very high statistics available from STAR that the calculation can be
considered in the range \( 15<Q_{3}<120 \)\,MeV/\( c \). This range
is large enough to provide reliable extrapolation to \( Q_{3}=0 \).

Data for the present results are from about $300$K
events taken during the
$\sqrt{s_{NN}}=130$ GeV Au+Au
run at STAR using the Time Projection Chamber (TPC) \cite{NIM}
as the primary tracking detector. In the discussion that follows,
all phase space cuts and experimental corrections are similar to the
two-pion HBT analysis \cite{STARHBT}. A set of multiplicity classes
was created by taking the 12\% most central for the high-multiplicity
class and the next 20\% most central for the mid-multiplicity class. 
For both multiplicity bins, tracks were
constrained to have \( p_{T} \) in the range 
\( 0.125<p_{T}<0.5 \)\,GeV/\( c \),
and pseudorapidity \( \left| \eta \right| <1.0 \).
A vertex cut was also applied to events such that the vertex
along the z-axis (beam direction) had to fall within $\pm75$ cm
of the center of the detector. 
In the range \( 15<Q_{3}<120 \)\,MeV/\( c \),
approximately 150 million triplets were included in both the negative
and positive pion studies.

The \( C_{2} \) correlation function was corrected for Coulomb repulsion
with a finite Gaussian source approximation, using an integration
of Coulomb wave functions \cite{Coulomb}. In calculating \( C_{3} \),
the correction was applied by taking the product of three two-pion
corrections, obtained from the three possible pairs formed from each
mixed-event
triplet. This type of correction approximates the three-body Coulomb
problem to first order \cite{TherMath,3PWaveFunction}. 
Other methods to more accurately estimate
the true three-body Coulomb effect show a 5-10\% smaller correction 
\cite{3-body-CC}. This difference was applied
to the Coulomb correction factor in calculating \( C_{3} \), and
the resulting shifts in the \( r_{3} \) function were found to be
within systematic uncertainties. A separate study
examined the effect of inappropriately applying the Coulomb correction
to pions which come from long-lived resonances \cite{scattres}. 
Using a rescattering
model \cite{TJH}, the value of \( r_{3} \) was found to increase 
by 10\% when pairs and triplets of pions
which contain pions from long-lived resonances were inappropriately
Coulomb corrected. Effects of finite momentum 
resolution on $r_3$ were also studied
using this model and were found to be insignificant. The $1\sigma$
uncertainty in determining $Q_3$ is found to be about $10$ MeV/c.

\begin{figure}
{\centering \resizebox*{\columnwidth}{!}{\includegraphics{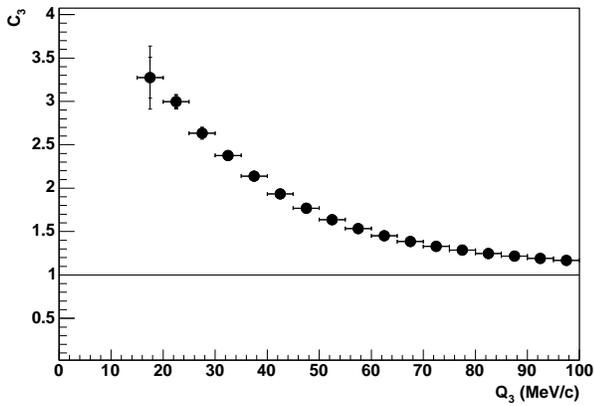}} \par}

\caption{\protect Three-pion correlation function for 
central Au-Au events using
\protect\( \pi ^{-}\protect \) triplets. Statistical and 
statistical+systematic
errors are shown.}

{\centering \label{FIG:ThreeParticleCF}\par}
\end{figure}

Figure \ref{FIG:ThreeParticleCF} shows the \( C_{3} \) correlation
function for negatively charged pions in the high-multiplicity bin.
The shape of $C_3$ is mostly built up of products of two-pion
correlations with the effect of true three-pion correlations being
more subtle. At large $Q_3$, $C_3$ approaches unity and for
an ideal pion source, i.e. $\lambda=1$, $C_3$ would approach
$6$ at $Q_3=0$ (this is not the present case since $\lambda<1$).
A Gaussian parameterization is inadequate to describe this
correlation function; this is consistent with results obtained in
other experiments and a simulation\cite{TJH,WA98,NA44}.
In calculating \( r_{3} \),
the actual binned values of the correlation function for the various
values of \( Q_{3} \) are used instead of a fit \cite{TJH}. In order
to use Eq.(~\ref{CosPhi}), triplets are obtained that pass all of
the momentum space and experimental cuts. \( Q_{3} \), \( Q_{12} \),
\( Q_{23} \) and \( Q_{31} \) are calculated from the triplet and
the three pairs that can be formed from the triplet. The values 
\( C_{3}\left( Q_{3}\right)  \),
\( C_{2}\left( Q_{12}\right)  \), \( C_{2}\left( Q_{23}\right)  \)
and \( C_{2}\left( Q_{31}\right)  \) are then computed 
from the binned actual
two- and three-pion correlation functions. These values are then used
to calculate \( r_{3} \), which is then binned as a function of \( Q_{3} \).
The average for each bin is then calculated to obtain the final result.
Systematic uncertainties are greatest at the low \( Q_{3} \) end due to
track merging effects and the uncertainty in the Coulomb correction. The
parameters controlling these effects were modified \(\pm 20\% \) from
the nominal values to obtain the overall systematic uncertainty in each
bin.
\begin{figure}
{\centering \resizebox*{0.9\columnwidth}{!}{\includegraphics{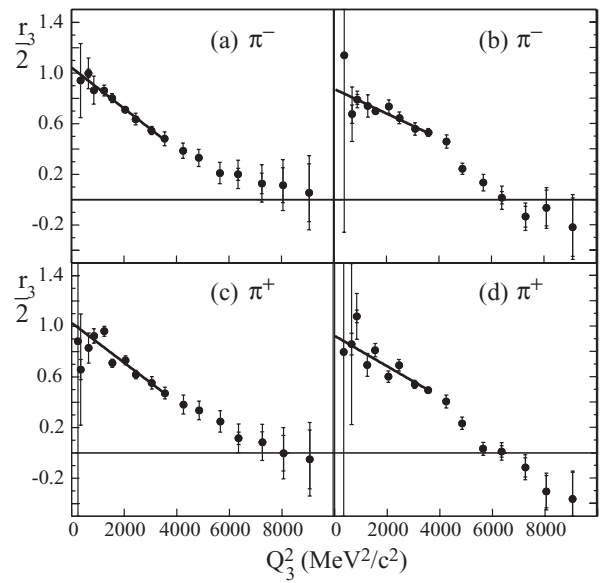}} \par}
\caption{\label{FIG:CosPhiBoth} Calculation of $r_3$
for (a) central and
(b) mid-central $\pi^{-}$ events,
and (c) central and (d) mid-central $\pi^{+}$ events.
The fits shown use Eq. (\ref{Factorization}).
Statistical and statistical+systematic uncertainties are shown.}
\end{figure}

The results for the two multiplicity bins are shown in Figure
\ref{FIG:CosPhiBoth} for
\( \pi ^{-} \) and \( \pi ^{+} \), plotted
as functions of \( Q_{3}^{2} \). Plotting in this way
is suggested by the theoretical
analysis in \cite{HZ} which shows that the leading relative momentum
dependencies in the numerator and denominator of Eq. (\ref{CosPhi})
are quadratic \cite{fn1}, allowing for a linear extrapolation of
the results shown in Figure~\ref{FIG:CosPhiBoth}
to \( Q_{3}{\, =\, }0 \) by fitting them to the form 
\begin{equation}
\label{Factorization}
r_{3}\left( Q_{3}\right)/2 =r_{3}(0)/2-\alpha\, Q^{2}_{3}.
\end{equation}
where $r_{3}(0)/2$ and $\alpha$ are fit parameters.
From Figure \ref{FIG:CosPhiBoth} it appears
that the normalized three-pion correlator \( r_{3}(Q_{3}) \) does
indeed show a leading quadratic dependence for the smaller $Q^{2}_{3}$
values (Eq. (\ref{Factorization}) was fit to the range $0<Q_3<60$ MeV/c).

\begin{figure}
{\centering \resizebox*{0.9\columnwidth}{!}{\includegraphics{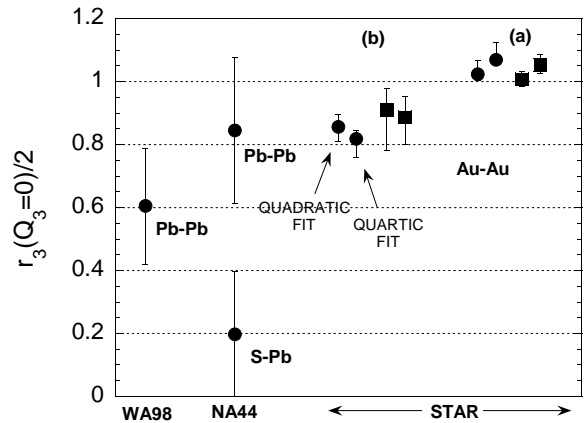}} \par}

\caption{$r_{3}(Q_3=0)/2$ from
STAR and two other experiments \cite{NA44,WA98}. For STAR,
(a) central and (b) mid-central results are shown for
\protect\( \pi ^{-}\protect \)
(circles) and \protect\( \pi ^{+}\protect \) (squares)
data. The other experiments use \protect\( \pi ^{-}\protect \) data
only.  STAR results for fitting with both a quadratic and quartic
functions are shown.}

{\centering \label{FIG:SummaryCosPhi}\par}
\end{figure}

The resulting intercepts \( r_{3}(0)/2 \) are shown in Figure
~\ref{FIG:SummaryCosPhi},
along with the results of WA98 and NA44.
Error bars for STAR points are statistical+systematic.
As mentioned earlier, the systematic error is computed by varying
several parameters independently, including particle track cut
parameters. The variation of the parameters is seen to produce,
in general, asymmetric variations in the extracted intercepts.
Intercepts from the quadratic fits as well as from quartic fits
(i.e. adding a quartic term to Eq. (3) and fitting over
the broader range $0<Q_3<120$ MeV/c)
are shown for comparison,
and are seen to agree within errors.
The STAR \( \pi ^{+} \) and \( \pi ^{-} \) results
are also seen to agree within error bars.
NA44 reported a result close
to unity for Pb-Pb interactions, but a much lower result for S-Pb
\cite{NA44}, both with no clear \( Q_{3} \) dependence.
The Pb-Pb result from WA98 is somewhat smaller than
that from NA44, although they agree within error bars,
and the \( Q_{3} \)-dependence in their result is similar to what
is seen in STAR \cite{WA98}.

\begin{figure}
{\centering \resizebox*{0.9\columnwidth}{!}{\includegraphics{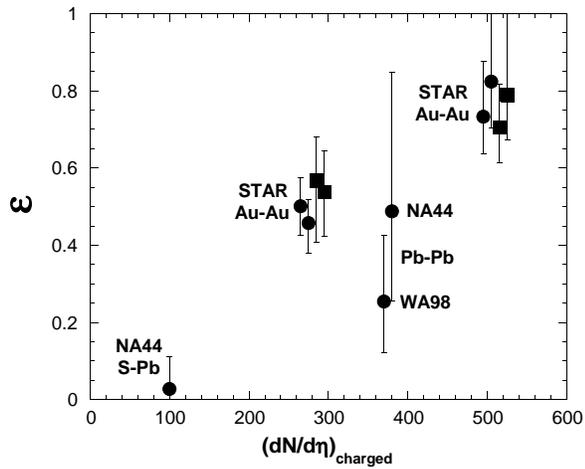}} \par}

\caption{\protect Chaotic fraction, calculated from Eq.
(\ref{Chaotic Fraction}),
and plotted versus charged particle multiplicity 
per unit pseudorapidity for the same 
experiments as in Figure \ref{FIG:SummaryCosPhi}. The meanings of the
symbols used in this figure are
the same as in Figure \ref{FIG:SummaryCosPhi}.
}

{\centering \label{FIG:SummaryCF}\par}
\end{figure}

Figure \ref{FIG:SummaryCF} shows the results from the 
calculation of \( \varepsilon  \)
for STAR's measurements, and for those from WA98 and NA44,
plotted versus charged particle multiplicity per unit
pseudorapidity, $dN/d\eta$.
The calculation was done starting with the results of
Figure \ref{FIG:SummaryCosPhi}, decreasing them by 10\% to
approximately take into
account the overcorrection produced by Coulomb-correcting long-lived
resonances (see earlier discussion) and using Eq. (2).
It was assumed that the 10\% correction also applies to the SPS
data, and to be conservative, a $\pm 5\%$ systematic uncertainty on the
correction (i.e. $10\% \pm 5\%$) was included in all of the error bars 
shown. The plot
shows an increasing trend in the STAR 
\( \pi ^{-} \) and \( \pi ^{+} \) results going
from mid-central to central collisions. For the mid-central
data, the results for \( \varepsilon  \) show
a partially chaotic source, as seen in the SPS results.
The central data appear to give a mostly chaotic pion source.
Including the SPS measurements into the overall systematics, there appears
to be, within the uncertainties shown, a systematic 
increase in \( \varepsilon  \) with increasing
$dN/d\eta$, the smallest value being for SPS S-Pb collisions
and the largest value for STAR central Au-Au collisions
($dN/d\eta$ for charged particles at mid-$\eta$ for SPS S-Pb,
SPS Pb-Pb, STAR mid-central, and STAR central are approximately
100(scaled from S-S), 370, 280, and 510, 
respectively\cite{na49mult,phobosmult}).
It is also found for the STAR results that the upper limit on the 
two-pion $\lambda$-parameter obtained from \( \varepsilon  \) using
the relationship mentioned earlier is in the range $0.71-0.81$ for
mid-central and $0.91-0.97$ for central events. The actual values for
$\lambda$ extracted from STAR $\pi^{-}-\pi^{-}$ HBT measurements are 
0.53$\pm$0.02
for mid-central and 0.50$\pm$0.01 for central events
(the $\pi^{+}-\pi^{+}$ values agree with these within errors)\cite{STARHBT}.
The lower $\lambda$-values extracted from the two-pion
experiment can be explained in terms of long-lived
resonance effects, which nicely cancel out in a three-pion
analysis\cite{TJH}.

In summary, we have presented three-pion HBT results for 
\( \sqrt{s_{NN}}{\, =\, }130 \)\,GeV
data at STAR, and have shown that for the central multiplicity class the
STAR data indicate a large degree of chaoticity in the source at freeze-out,
whereas for the mid-central class the source is less chaotic.
Our \( r_{3} \) results are close to those extracted in SPS Pb+Pb
collisions, but differ from the low value obtained in SPS S+Pb collisions.
The comparison between SPS and STAR results suggests a systematic
dependence of the chaoticity on particle multiplicity.
High statistics from STAR have allowed a normalized three-pion correlator
calculation that extends to \( 120 \) MeV/\( c \) in \( Q_{3} \),
and the dependence on this variable has been shown to be quadratic
in nature for low \( Q_{3} \).
STAR's measured values provide increased confidence in the validity
of standard HBT analyses based on the assumption of a chaotic source
for central collisions at RHIC.

\begin{acknowledgments}
We wish to thank the RHIC Operations Group and the RHIC Computing Facility
at Brookhaven National Laboratory, and the National Energy Research
Scientific Computing Center at Lawrence Berkeley National Laboratory
for their support. This work was supported by the Division of Nuclear
Physics and the Division of High Energy Physics of the Office of Science of
the U.S. Department of Energy, the United States National Science Foundation,
the Bundesministerium fuer Bildung und Forschung of Germany,
the Institut National de la Physique Nucleaire et de la Physique
des Particules of France, the United Kingdom Engineering and Physical
Sciences Research Council, Fundacao de Amparo a Pesquisa do Estado de Sao
Paulo, Brazil, the Russian Ministry of Science and Technology, the
Ministry of Education of China, the National Natural Science Foundation
of China, the Swiss National Science Foundation, and the Grant Agency
of the Czech Republic.
\end{acknowledgments}

\end{document}